\documentclass[journal,onecolumn]{IEEEtran}
\IEEEoverridecommandlockouts

\usepackage{hyperref}

\ifCLASSINFOpdf
  \usepackage{graphicx}

  \graphicspath{{../pdf/}{../jpeg/}}

  \DeclareGraphicsExtensions{.pdf,.jpeg,.png}
\else

\fi

\usepackage{makecell}

\usepackage{amsmath}

\usepackage{bbold}

\usepackage{xcolor}

\usepackage{titlesec}
\titlespacing\section{0pt}{3pt plus 1pt minus 1pt}{2pt plus 1pt minus 1pt}
\titlespacing\subsection{0pt}{2pt plus 1pt minus 2pt}{2pt plus 0pt minus 1pt}

\begin{document}

\title{Evaluating Encoding Strategies for Closed-Loop Classification in Biological Neural Networks}

\author{
    \IEEEauthorblockN{
    Martin Schottlender, Veronika Volkova, Pengjie Zhou, Ruifeng Zheng, Frank H.\,P. Fitzek and Pit Hofmann\\
    }
    \thanks{M.~Schottlender, V.~Volkova, P.~Zhou, R.~Zheng, F. H.\,P.~Fitzek, and P.~Hofmann are with the Deutsche Telekom Chair of Communication Networks, Dresden University of Technology, Germany; F.~Fitzek, and P.~Hofmann are also with the Centre for Tactile Internet with Human-in-the-Loop (CeTI), Dresden, Germany, Emails: \{martin.schottlender, veronika.volkova, pengjie.zhou, ruifeng.zheng,  frank.fitzek, pit.hofmann\}@tu-dresden.de. This work has been submitted to the IEEE for possible publication. Copyright may be transferred without notice, after which this version may no longer be accessible.
}}

\maketitle

\begin{abstract}
 Interfacing with Biological Neural Networks (BNNs) requires encoding information into stimulation patterns that can be effectively processed and that enable the underlying system to adapt. Nevertheless, the role of stimulation encoding remains poorly understood. In this work, we compare multiple encoding strategies, including rate-based, phase-based, burst-based, and time-to-first-spike temporal encodings, in a closed-loop neural classification task using cultured BNNs. We encode visual inputs as spatiotemporal stimulation patterns delivered via a Multi-Electrode Array (MEA) and evaluate classification performance for each encoding scheme. We find that burst-based temporal encoding yields the highest observed performance, achieving up to $95.6\ \%$ accuracy in a binary classification task, compared to substantially lower performance from rate- and phase-based approaches.  We further show that performance is highly sensitive to the spatial distribution of stimulation, with suboptimal electrode selection significantly degrading accuracy. These findings indicate that effective interfacing with biological neural systems requires the joint optimization of temporal and spatial encoding strategies, and highlight temporal encoding as a key design dimension for bio-digital computing.
\end{abstract}

\begin{IEEEkeywords}
Synthetic Biological Intelligence, Biological Neural Networks, Biocomputing, Closed-loop Systems.
\end{IEEEkeywords}

\IEEEpeerreviewmaketitle

\section{Introduction}
\IEEEPARstart{B}{iologically-inspired} computing, a paradigm in which biological systems inspire the function of computational systems, has been increasingly studied by the scientific community due to its inherent energy efficiency. For example, the human brain is six orders of magnitude more energy-efficient than supercomputers, matching its computing power \cite{smirnovaOrganoidIntelligenceOI2023}. The use of actual BNNs in hybrid bio-digital engineered systems has recently sparked interest in both the computer science and the neuroscience community \cite{schottlenderSyntheticBiologicalIntelligence2026}. This field has been coined \textit{Synthetic Biological Intelligence} (SBI) \cite{kaganHarnessingIntelligenceBrain2025}.

SBI has been made available with the development of induced Pluripotent Stem Cells (iPSCs) and MEAs \cite{smirnovaOrganoidIntelligenceOI2023}. Biological neurons, in particular human cortical neurons, are very complex units that offer plasticity and adaptability, and provide emergent properties when stimulating them as part of a population \cite{kaganVitroNeuronsLearn2022}. Recently, commercially available platforms such as Cortical Labs' CL-1 and Finalspark's Neuroplatform have enabled researchers to access this technology through Application Programming Interfaces (APIs) and practical Graphical User Interfaces (GUIs) \cite{kaganCL1PlatformTechnology2025,jordanOpenRemotelyAccessible2024}.

A central open question in SBI is still how information should be encoded into stimulation patterns so that biological neural cultures can produce discriminable and feedback-sensitive responses. Existing closed-loop BNN studies have demonstrated adaptive behavior, but the roles of stimulation encoding, decoding representation, and electrode population selection remain insufficiently characterized \cite{chenOverviewVitroBiological2023}. This is particularly important for the standardization and reproducibility of experiments, since the same temporal stimulation strategy may produce different responses depending on electrode activity, drift, and spatial organization.

To address this gap, we evaluate several temporal encoding and decoding strategies in a common closed-loop classification protocol on the Cortical Labs CL-1 platform. Visual inputs from a two-symbol MNIST task are mapped to electrical stimulation patterns, neural responses are decoded into class labels, and reward- or punishment-based feedback is delivered to the culture. We then compare temporal encoding strategies and population electrode configurations to determine which design choices improve closed-loop classification performance.

Each encoding and decoding scheme evaluated in this work is derived from established neural coding and spike-train analysis methods, either already used for other BNN experiments or for neuromorphic computing \cite{guoNeuralCodingSpiking2021}. The main contributions of this paper are summarized as follows: (i) a closed-loop CL-1 experimental protocol comparing encoding and decoding strategies for a common classification task, (ii) an empirical within-culture evaluation showing the configuration with the best performance, and (iii) an analysis of population encoding showing the techniques used to optimize the performance for the given substrate. We also suggest future strategies for improving these frameworks and describe challenges when replicating these experiments.

\section{Bio-Digital System using the CL-1} \label{sec:bio_digital_system}

The standardization of engineered systems using BNNs for adaptive behavior is a key driver towards the availability of SBI \cite{schottlenderSyntheticBiologicalIntelligence2026}. However, this technology has demanding requirements, such as access to a wet lab to generate neural cultures from iPSCs and maintain them in the MEA.

\subsection{Cortical Cloud} \label{sec:cl1}

The Cortical Cloud from Cortical Labs provides real-time API access to their CL-1 platforms located in Melbourne, Australia \cite{hoganCLAPIRealTime2026}. The CL-1 provides a bio-digital interface to a neural substrate with an MEA, enabling both electrical stimulation and recording of neural activity. CL-1 cultures comprise approximately $8\cdot10^5$ human cortical neurons derived from iPSCs. The neural cultures are connected with an $8\, \mathrm{mm}^2$ two-dimensional MEA containing 64 electrodes, of which 60 are interactive.

The Cortical Cloud's GUI and API enable the user to define stimulation sequences and receive spike data streams with precise temporal resolution. Each stimulation is delivered as a sequence of electrical pulses parameterized by amplitude, frequency, and duration, which can be modulated according to the chosen encoding strategy.

\begin{figure}
  \centering
  \includegraphics[width=0.8\linewidth]{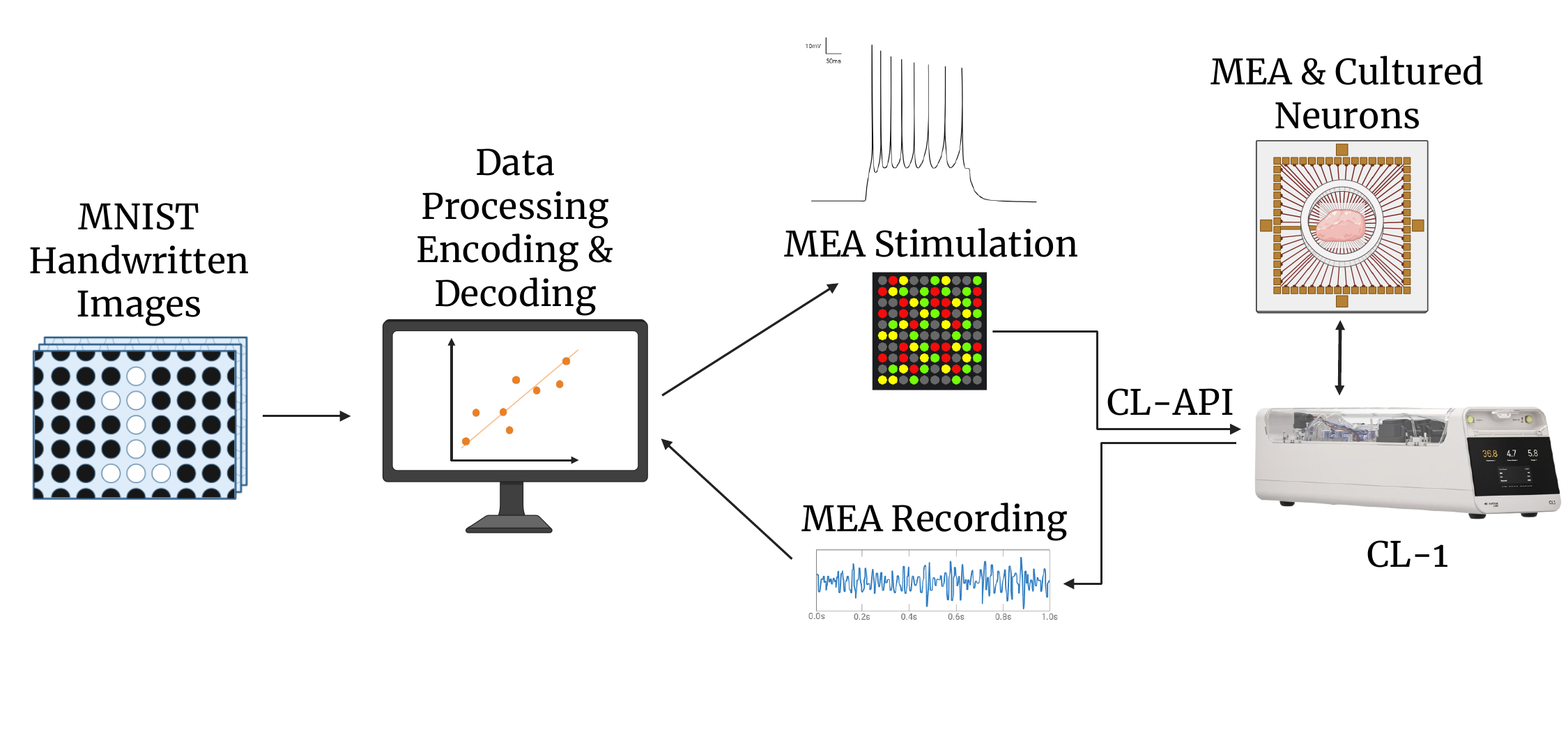}
  \caption{Schematic of the closed-loop CL-1 classification pipeline. MNIST symbols are converted into stimulation patterns using a selected spatial and temporal encoding strategy. The stimulation strategies are sent via an API to a CL-1, which provides the stimulation to the biological neural substrate through the MEA. The recorded spike responses are then decoded, and, based on the classification result, reward or punishment feedback is provided. Created in BioRender. Schottlender, M. (2026) \url{https://BioRender.com/d0x4giw}}
  \label{fig:cl1_system_schematic}
\end{figure}

The neural responses are recorded as discrete spike events, each characterized by an electrode index and a timestamp. These spike trains form the basis for the decoding process described in Section \ref{sec:enc_dec}. To ensure consistent analysis, responses are collected within a fixed time window following the end of stimulation.

\subsection{MNIST Classification Task and Closed-Loop System}
\label{subsec:mnist_closed_loop}

To evaluate the effectiveness of different encoding strategies, we employ a classification task based on the MNIST dataset of handwritten digits \cite{dengMNISTDatabaseHandwritten2012,gressmannLearningHarnessInVitro2025}. Each input image is converted into a spatiotemporal stimulation pattern, where pixel intensities are mapped to electrical stimulation parameters. For each symbol, a stimulation pattern is delivered through the MEA, and the resulting neural response is recorded. The recorded activity is then processed and decoded into a predicted label. This framework is illustrated in Fig. \ref{fig:cl1_system_schematic}.

The system operates in two phases: an open-loop phase and a closed-loop phase. The open-loop phase is necessary to establish a mapping between neural responses and task labels \cite{pimashkinAdaptiveEnhancementLearning2013,kuykendalClosedLoopCharacterizationNeuronal2017,robbinsGoalDirectedLearningCortical2024}, in which stimulation patterns are applied without feedback. Responses to known stimuli are first collected and used to train a decoder. This enables the transformation of neural activity into predicted labels during the closed-loop phase.

In the closed-loop setting, the neural culture receives feedback based on the decoding result. In each iteration $t$, an input symbol $x_t$ is encoded in an electrical stimulation pattern and sent to the neural culture. The evoked response is recorded over a fixed response window and transformed into a feature vector using the selected decoding representation. The classifier trained during the open-loop phase then predicts a label $\hat{y}_t$, which is compared with the ground-truth label $y_t$, and the outcome determines the feedback delivered to the neural culture.

This can be formalized with a binary reward signal: correct classifications receive positive reinforcement, while incorrect classifications receive negative reinforcement. Although very similar to Reinforcement Learning (RL) approaches, the reward and punishment signals are not used to update the decoder parameters online. Instead, the closed-loop feedback is intended to achieve adaptation within the substrate \cite{kaganVitroNeuronsLearn2022}.

The reward and punishment parameters can be seen in Tab. \ref{tab:reward_penalty}. These were selected based on \cite{kaganVitroNeuronsLearn2022}, considering the \textit{Free Energy Principle} \cite{isomuraCulturedCorticalNeurons2015,isomuraVitroNeuralNetworks2018}, in which it is assumed that neural populations aim to reduce uncertainty. Therefore, a ``predictable'' pattern, more tuned to the frequency neurons tend to operate, is used as a reward. Likewise, an ``unpredictable'' pattern, with a very low frequency in relation to the typical neural frequency, is selected as a punishment. The aim is for repeated exposure to this stimulus--response--feedback loop to promote the emergence of more discriminable and task-relevant neural activity over time.

\begin{table}[!t]
\renewcommand{\arraystretch}{1.3}
\caption{Specifications of the reward and punishment delivered to the neurons in the feedback loop.}
\label{tab:reward_penalty}
\centering
\begin{tabular}{|c|c|c|}
\hline
Parameter & Positive feedback & Negative feedback \\
\hline
Stimulation amplitude & $2.5\ \mu \mathrm{A}$ & $2.5\ \mu \mathrm{A}$ \\
\hline
Stimulation pulse width & $160\ \mu s$ & $160\ \mu s$ \\
\hline
Stimulation frequency & $100\ \mathrm{Hz}$ & $5\ \mathrm{Hz}$ \\
\hline
Duration & $0.1\ \mathrm{s}$ & $4\ \mathrm{s}$ \\
\hline
\end{tabular}
\vspace{-0.4cm}
\end{table}

\section{Encoding and Decoding Methods for BNNs}
\label{sec:enc_dec}

The most suitable encoding and decoding methods might vary by task, and their specific parameters also depend on the characteristics of the neural culture itself \cite{schottlenderSyntheticBiologicalIntelligence2026}.

We compare four temporal encoding strategies. Rate encoding maps input intensity to stimulation frequency, a very conventional neural coding approach \cite{guoNeuralCodingSpiking2021}. Burst encoding maps the input intensity to the inter-spike interval of short pulse trains, motivated by bursts in in vivo and in vitro neural networks \cite{bakkumRemovingAIEmbodied2004}. Time-to-first-spike (TTFS) encoding maps input intensity to stimulation latency, such that stronger inputs are delivered earlier \cite{bakkumLongTermActivityDependentPlasticity2008}. Phase encoding represents input intensity as a phase offset relative to an oscillatory stimulation cycle, inspired by phase-based neural coding and useful for synchronization between neural cultures \cite{kayserSpikePhaseCodingBoosts2009}.

\begin{figure}
 \centering
 \includegraphics[width=0.6\linewidth]{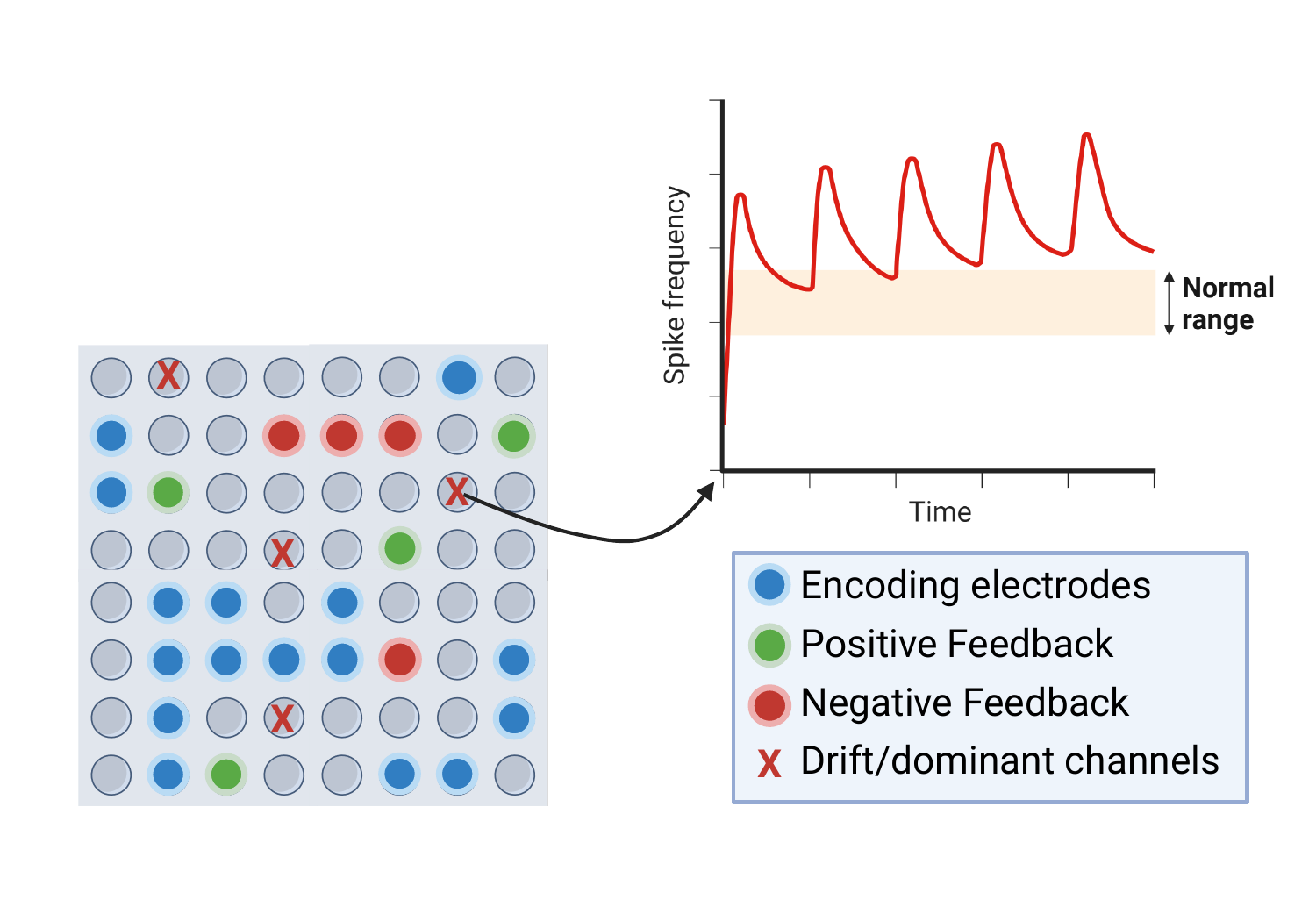}
 \caption{Population encoding protocol for the CL-1 MEA. Encoding electrodes were selected from moderately active channels. Feedback electrodes were selected from strong but non-dominant channels. Drift-affected or dominant electrodes were excluded from the optimized configuration. Created in BioRender. Schottlender, M. (2026) \url{https://BioRender.com/mbkjugz}}
 \label{fig:mea_population_strategy}
\end{figure}

In addition to rate and temporal encoding mechanisms, we assess population encoding methods that determine how stimulation electrodes are selected and assigned to input and feedback signals. For MEA-based cultures, electrode selection is important because activity may be unevenly distributed across the substrate, and some electrodes may exhibit drift, noise, or dominant responses. We show a schematic of the protocol we applied in Fig. \ref{fig:mea_population_strategy}. The feedback electrodes were selected to be the most active, but not overly dominant, according to the number of spikes detected in a baseline period. Moderately active channels were chosen for the stimulation of the input. We compare an optimized population configuration with a drift-affected configuration, using activity level, response stability, spatial distribution, and dominance as selection criteria \cite{bisioClosedLoopSystemsVitro2019}.

Decoding methods transform recorded neural responses into an interpretable representation for the task at hand. This requires identifying spikes on each electrode, a process called \textit{spike sorting}, and extracting the temporal and spatial features of the spike trains \cite{buccinoSpikeSortingNew2022}. We evaluate two decoding methods. TTFS decoding assigns labels based on the earliest evoked response across recording electrodes, while binned raster decoding converts spike trains into spike-count vectors over fixed temporal bins \cite{huaMicroelectrodeArraysCultured2025,guoNeuralCodingSpiking2021}. These representations capture complementary timing- and rate-based response features and are commonly used in spike-train analysis pipelines.

\section{Results}
\label{results}

We carried out the experiments described in Section \ref{subsec:mnist_closed_loop} on a single CL-1 biological neural culture with the parameters reported in Section \ref{sec:cl1}. We used a binary MNIST task with digits 0 and 1 as the two target symbols. The images were downsampled from 28x28 pixels to 4x4 by average pooling to match 16 encoding electrodes. Each closed-loop session consisted of 1000 symbols and required approximately three hours, followed by a three hour resting period. As the availability of biological substrates is highly constrained, the number of sessions and their parameters were unequal across encoding families. Therefore, the reported accuracies should be interpreted as evidence of encoding-dependent performance within one biological substrate rather than population-level estimates across cultures. Future work will replicate these experiments in other cultures, both with the CL-1 and other platforms, to analyze inter-substrate variability.

Closed-loop performance is tracked through the running classification accuracy over all completed trials:

\begin{equation}
    \mathrm{Acc}_\mathrm{N} = \frac{1}{\mathrm{N}}\sum_{n=1}^{\mathrm{N}} \mathbb{1}(\hat{y}_n = y_n),
\end{equation}

where $\mathbb{1}(\cdot)$ is the indicator function and $\mathrm{N}$ is the number of decoded trials. 

Since several sessions for each encoding--decoding family correspond to parameter variants rather than exact repetitions, Tab.~\ref{tab:encoding_results} reports the best observed session for each family together with the number of screening sessions. Four-symbol sessions were excluded from this comparison. Burst-based encoding with TTFS decoding achieves the highest classification performance, reaching up to 95.6\% closed-loop accuracy in the binary classification task. In contrast, rate-based and phase-based encoding strategies produced lower performances, particularly in the closed-loop setting. This indicates that the temporal structure of stimulation plays a critical role in shaping effective neural responses.

\begin{table}[!t]
\renewcommand{\arraystretch}{1.3}
\caption{Best observed classification accuracy for different encoding--decoding methods in the two-symbol closed-loop MNIST task. Sessions include parameter variants within the same encoding--decoding family.}
\label{tab:encoding_results}
\centering
\begin{tabular}{|c|c|c|c|c|}
\hline
\makecell{Encoding\\Method} & Decoder & Sessions & \makecell{Open-Loop\\(\%)}  & \makecell{Closed-Loop\\(\%)} \\
\hline
Burst  & TTFS & 55 & \textbf{91.9} & \textbf{95.6} \\
Rate   & Binned Raster & 3 & 85.9 & 62.1 \\
Rate & TTFS  & 7 & 67.2 & 62.8 \\
Phase & Binned Raster  & 2 & 72.7 & 52.5 \\
TTFS & TTFS & 1  & 67.8 & 51.0 \\
\hline
\end{tabular}
\end{table}

\begin{figure}
  \centering
  \includegraphics[width=0.5\linewidth]{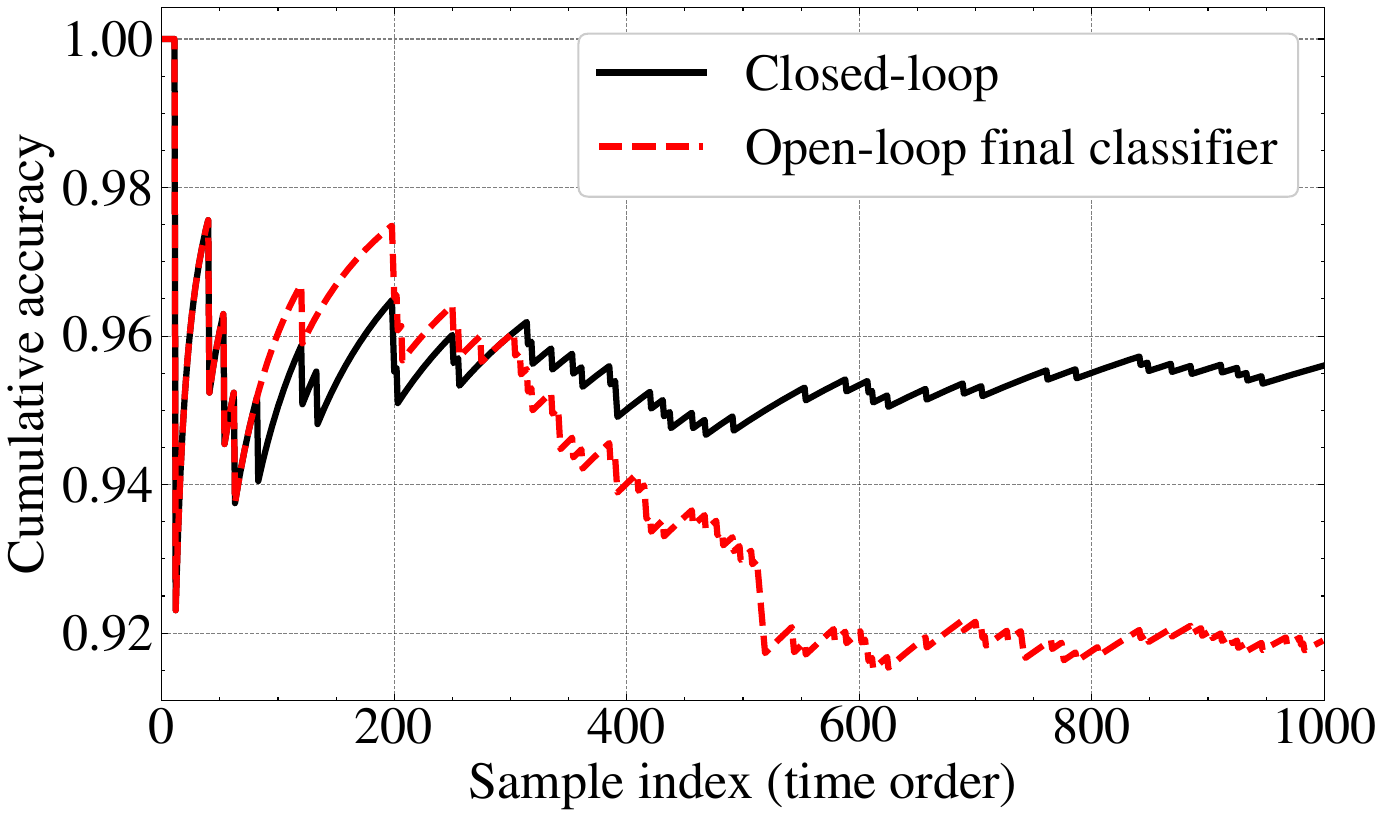}
  \caption{Cumulative accuracy over time for an example of the best encoding-decoding combination for MNIST with two symbols.} \label{fig:accuracy_time}
  \vspace{-0.4cm}
\end{figure}

A notable observation is the discrepancy between open-loop and closed-loop performance across encoding methods. Most configurations achieve relatively strong open-loop accuracy, with a significantly lower closed-loop performance. Only the best-performing Burst+TTFS configuration improved from open-loop to closed-loop evaluation. An example of how accuracy develops over time is shown in Fig. \ref{fig:accuracy_time}. This suggests that adaptive behavior depends on using stimulation patterns that better match the response dynamics of the neuronal population. It also highlights the importance of feedback dynamics, suggesting that encoding strategies must not only produce separable responses in the open loop but also interact effectively with the reward-based feedback mechanism. 

The impact of population encoding and spatial stimulation is further explored in Tab. \ref{tab:population_effect}. The performance of burst-based methods is strongly dependent on the population encoding strategy. Electrodes with a strong but not dominant electrical response were chosen for the feedback, while moderately strong ones were used for the encoding. Drifting electrodes or highly dominant channels led to a substantial reduction in classification accuracy. Furthermore, using 16 channels for reward and 16 channels for punishment reduced closed-loop performance below the open-loop baseline, suggesting that overly distributed feedback may reduce the specificity of the reinforcement signal and disrupt discriminable response patterns. These results indicate that balanced electrode selection is necessary to preserve response diversity while avoiding redundancy. The optimal configuration, consisting of 16 encoding electrodes and four electrodes per feedback signal, achieved the highest performance.

\begin{table}[!t]
\renewcommand{\arraystretch}{1.3}
\caption{Impact of population encoding strategy on burst-based methods. Sessions include parameter variants within the same encoding--decoding family.}
\label{tab:population_effect}
\centering
\begin{tabular}{|c|c|c|c|c|c|}
\hline
\makecell{Population\\Encoding} & \makecell{Encoding\\electrodes} & \makecell{Feedback\\electrodes} & Sessions & \makecell{OL\\(\%)} & \makecell{CL\\(\%)} \\
\hline
Optimized & 16 & 4+4 & 4 & \textbf{91.9} & \textbf{95.6} \\
\makecell{Distributed\\feedback} & 16  & 16+16 & 12 &  86.9 & 87.6 \\
Drift-Affected & 16 & 16+16  & 39 &  88.0 & 83.3 \\
\hline
\end{tabular}
\vspace{-0.4cm}
\end{table}

\begin{table}[!t]
\renewcommand{\arraystretch}{1.3}
\caption{Parameter configuration for the best-performing experiment with burst encoding + TTFS decoding.}
\label{tab:burst_ttfs_params}
\centering
\begin{tabular}{|c|c|c|c|c|}
\hline
Wait (s)  & $f_{\min}$ (Hz) & $f_{\max}$ (Hz) & $N_S$ & Input threshold\\
 \hline
 0.01 & 8.33 & 50 & 5 & 0.08 \\
\hline
\end{tabular}
\vspace{-0.6cm}
\end{table}

When the task was increased from two to four classes, the closed-loop accuracy decreased from $95.6\%$ to $50.6\%$, suggesting that scalability remains limited under the current stimulation and decoding framework. This trend indicates that, under the current stimulation and decoding framework, the representational capacity of the neural culture may be insufficient to support more complex classification tasks without further optimization.

These results suggest that both temporal encoding and spatial electrode selection are critical factors in determining performance. The interaction between these two dimensions suggests that encoding strategies cannot be evaluated in isolation, as suboptimal population configurations can mask the advantages of otherwise effective temporal encoding methods.

\section{Conclusion and Future Work}

In this work, we evaluated the impact of temporal and population encoding strategies for closed-loop classification in biological neural networks. Using a real-time bio-digital interface on the Cortical Labs CL-1 platform, we observed that burst-based temporal encoding, particularly when combined with TTFS decoding, outperforms rate and phase-based approaches.

This indicates that effective interfacing with BNNs depends not only on temporal encoding but also critically on spatial stimulation strategies. We show that suboptimal population encoding, including the use of drifting or highly dominant electrodes, can substantially degrade performance. This suggests that the temporal and spatial encoding dimensions are strongly coupled and should be optimized jointly.

Although high accuracy can be achieved in binary classification, performance decreases as the number of classes increases, suggesting limitations in the representational capacity of the neural culture under the current framework. Research on organoid topography suggests that modular approaches have the potential to improve the performance of high-dimensional class differentiation problems \cite{levyEnhancementNeuralRepresentation2012,isomuraCulturedCorticalNeurons2015,forroModularMicrostructureDesign2018}. 

In later experiments, neural activity was observed to be strongly concentrated in a localized region of the MEA. This degraded the classification performance even after using optimization techniques. This suggests that high activity alone is not sufficient for effective population encoding and that excessively localized or dominant activity can reduce the diversity and discriminability of neural responses. This highlights substrate monitoring as an important research direction for improving the reproducibility of closed-loop bio-digital systems and the development of mitigation techniques.

These findings contribute to the development of standardized encoding and decoding principles for biological neural interfaces, an essential step to advance SBI systems. Future work will focus on improving decoding strategies through more advanced machine learning models, exploring alternative stimulation modalities beyond electrical signals, and investigating structured or modular neural topologies to enhance scalability.

\bibliographystyle{IEEEtran}
\bibliography{references}

\begin{thebibliography}{10}
\providecommand{\url}[1]{#1}
\csname url@samestyle\endcsname
\providecommand{\newblock}{\relax}
\providecommand{\bibinfo}[2]{#2}
\providecommand{\BIBentrySTDinterwordspacing}{\spaceskip=0pt\relax}
\providecommand{\BIBentryALTinterwordstretchfactor}{4}
\providecommand{\BIBentryALTinterwordspacing}{\spaceskip=\fontdimen2\font plus
\BIBentryALTinterwordstretchfactor\fontdimen3\font minus \fontdimen4\font\relax}
\providecommand{\BIBforeignlanguage}[2]{{%
\expandafter\ifx\csname l@#1\endcsname\relax
\typeout{** WARNING: IEEEtran.bst: No hyphenation pattern has been}%
\typeout{** loaded for the language `#1'. Using the pattern for}%
\typeout{** the default language instead.}%
\else
\language=\csname l@#1\endcsname
\fi
#2}}
\providecommand{\BIBdecl}{\relax}
\BIBdecl

\bibitem{smirnovaOrganoidIntelligenceOI2023}
L.~Smirnova \emph{et~al.}, ``Organoid {Intelligence} ({OI}): {The} {New} {Frontier} in {Biocomputing} and {Intelligence}-in-a-{Dish},'' \emph{Front Sci}, vol.~1, Feb. 2023.

\bibitem{schottlenderSyntheticBiologicalIntelligence2026}
M.~Schottlender \emph{et~al.}, ``\BIBforeignlanguage{en}{Synthetic {Biological} {Intelligence}: {System}-{Level} {Abstractions} and {Adaptive} {Bio}-{Digital} {Interaction}},'' \emph{\BIBforeignlanguage{en}{arXiv}}, Apr. 2026, {Art. no. 2604.27933}.

\bibitem{kaganHarnessingIntelligenceBrain2025}
B.~J. Kagan \emph{et~al.}, ``\BIBforeignlanguage{en}{Harnessing {Intelligence} from {Brain} {Cells} {In} {Vitro}},'' \emph{\BIBforeignlanguage{en}{Neuroscientist}}, vol.~31, no.~5, pp. 536--555, Oct. 2025.

\bibitem{kaganVitroNeuronsLearn2022}
------, ``\BIBforeignlanguage{en}{In {Vitro} {Neurons} {Learn} and {Exhibit} {Sentience} {When} {Embodied} in a {Simulated} {Game}-{World}},'' \emph{\BIBforeignlanguage{en}{Neuron}}, vol. 110, no.~23, pp. 3952--3969, Dec. 2022.

\bibitem{kaganCL1PlatformTechnology2025}
B.~J. Kagan, ``\BIBforeignlanguage{en}{The {CL1} as a {Platform} {Technology} to {Leverage} {Biological} {Neural} {System} {Functions}},'' \emph{\BIBforeignlanguage{en}{Nat. Rev. Bioeng}}, vol.~3, no.~9, pp. 724--725, Sep. 2025.

\bibitem{jordanOpenRemotelyAccessible2024}
F.~D. Jordan, M.~Kutter, J.-M. Comby, F.~Brozzi, and E.~Kurtys, ``Open and {Remotely} {Accessible} {Neuroplatform} for {Research} in {Wetware} {Computing},'' \emph{Front. Artif. Intell}, vol.~7, May 2024.

\bibitem{chenOverviewVitroBiological2023}
Z.~Chen \emph{et~al.}, ``\BIBforeignlanguage{en}{An {Overview} of {In} {Vitro} {Biological} {Neural} {Networks} for {Robot} {Intelligence}},'' \emph{\BIBforeignlanguage{en}{Cyborg Bionic Syst}}, vol.~4, Jan. 2023, {Art. no. 0001}.

\bibitem{guoNeuralCodingSpiking2021}
W.~Guo, M.~E. Fouda, A.~M. Eltawil, and K.~N. Salama, ``\BIBforeignlanguage{English}{Neural {Coding} in {Spiking} {Neural} {Networks}: {A} {Comparative} {Study} for {Robust} {Neuromorphic} {Systems}},'' \emph{\BIBforeignlanguage{English}{Front. Neurosci}}, vol.~15, Mar. 2021.

\bibitem{hoganCLAPIRealTime2026}
D.~Hogan \emph{et~al.}, ``{CL} {API}: {Real}-{Time} {Closed}-{Loop} {Interactions} with {Biological} {Neural} {Networks},'' \emph{arXiv}, Feb. 2026, {Art. no. 2602.11632}.

\bibitem{dengMNISTDatabaseHandwritten2012}
L.~Deng, ``The {MNIST} {Database} of {Handwritten} {Digit} {Images} for {Machine} {Learning} {Research},'' \emph{IEEE Signal Process. Mag}, vol.~29, no.~6, pp. 141--142, Nov. 2012.

\bibitem{gressmannLearningHarnessInVitro2025}
F.~Gressmann and L.~Rauchwerger, ``\BIBforeignlanguage{en}{Learning to {Harness} {In}-{Vitro} {Biological} {Neural} {Networks}},'' in \emph{\BIBforeignlanguage{en}{Principles and {Practices} of {Building} {Parallel} {Software}}}.\hskip 1em plus 0.5em minus 0.4em\relax Springer Nature Switzerland, Jun. 2025, vol. 14564, pp. 78--89.

\bibitem{pimashkinAdaptiveEnhancementLearning2013}
A.~Pimashkin, A.~Gladkov, I.~Mukhina, and V.~Kazantsev, ``\BIBforeignlanguage{English}{Adaptive {Enhancement} of {Learning} {Protocol} in {Hippocampal} {Cultured} {Networks} {Grown} on {Multielectrode} {Arrays}},'' \emph{\BIBforeignlanguage{English}{Front. Neural Circuits}}, vol.~7, May 2013.

\bibitem{kuykendalClosedLoopCharacterizationNeuronal2017}
M.~L. Kuykendal, G.~S. Guvanasen, S.~M. Potter, M.~A. Grover, and S.~P. DeWeerth, ``\BIBforeignlanguage{en}{Closed-{Loop} {Characterization} of {Neuronal} {Activation} {Using} {Electrical} {Stimulation} and {Optical} {Imaging}},'' \emph{\BIBforeignlanguage{en}{Processes}}, vol.~5, no.~2, Jun. 2017.

\bibitem{robbinsGoalDirectedLearningCortical2024}
A.~Robbins \emph{et~al.}, ``\BIBforeignlanguage{en}{Goal-{Directed} {Learning} in {Cortical} {Organoids}},'' \emph{\BIBforeignlanguage{en}{bioRxiv}}, Dec. 2024.

\bibitem{isomuraCulturedCorticalNeurons2015}
T.~Isomura, K.~Kotani, and Y.~Jimbo, ``Cultured {Cortical} {Neurons} {Can} {Perform} {Blind} {Source} {Separation} {According} to the {Free}-{Energy} {Principle},'' \emph{PLoS Comput. Biol}, vol.~11, no.~12, Dec. 2015.

\bibitem{isomuraVitroNeuralNetworks2018}
T.~Isomura and K.~Friston, ``\BIBforeignlanguage{en}{In {Vitro} {Neural} {Networks} {Minimise} {Variational} {Free} {Energy}},'' \emph{\BIBforeignlanguage{en}{Sci. Rep}}, vol.~8, no.~1, Nov. 2018.

\bibitem{bakkumRemovingAIEmbodied2004}
D.~J. Bakkum \emph{et~al.}, ``\BIBforeignlanguage{en}{Removing {Some} ‘{A}’ from {AI}: {Embodied} {Cultured} {Networks}},'' in \emph{\BIBforeignlanguage{en}{Embodied {Artificial} {Intelligence}: {International} {Seminar}, {Dagstuhl} {Castle}, {Germany}, {July} 7-11, 2003. {Revised} {Papers}}}.\hskip 1em plus 0.5em minus 0.4em\relax Berlin, Heidelberg: Springer, Jul. 2004, pp. 130--145.

\bibitem{bakkumLongTermActivityDependentPlasticity2008}
D.~J. Bakkum, Z.~C. Chao, and S.~M. Potter, ``\BIBforeignlanguage{en}{Long-{Term} {Activity}-{Dependent} {Plasticity} of {Action} {Potential} {Propagation} {Delay} and {Amplitude} in {Cortical} {Networks}},'' \emph{\BIBforeignlanguage{en}{PLoS ONE}}, vol.~3, no.~5, May 2008.

\bibitem{kayserSpikePhaseCodingBoosts2009}
C.~Kayser, M.~A. Montemurro, N.~K. Logothetis, and S.~Panzeri, ``Spike-{Phase} {Coding} {Boosts} and {Stabilizes} {Information} {Carried} by {Spatial} and {Temporal} {Spike} {Patterns},'' \emph{Neuron}, vol.~61, no.~4, pp. 597--608, Feb. 2009.

\bibitem{bisioClosedLoopSystemsVitro2019}
M.~Bisio \emph{et~al.}, ``\BIBforeignlanguage{en}{Closed-{Loop} {Systems} and {In} {Vitro} {Neuronal} {Cultures}: {Overview} and {Applications}},'' in \emph{\BIBforeignlanguage{en}{In {Vitro} {Neuronal} {Networks}}}.\hskip 1em plus 0.5em minus 0.4em\relax Cham: Springer International Publishing, May 2019, vol.~22, pp. 351--387.

\bibitem{buccinoSpikeSortingNew2022}
A.~P. Buccino, S.~Garcia, and P.~Yger, ``\BIBforeignlanguage{en}{Spike {Sorting}: {New} {Trends} and {Challenges} of the {Era} of {High}-{Density} {Probes}},'' \emph{\BIBforeignlanguage{en}{Prog. Biomed. Eng}}, vol.~4, no.~2, May 2022.

\bibitem{huaMicroelectrodeArraysCultured2025}
S.~Hua \emph{et~al.}, ``\BIBforeignlanguage{en}{Microelectrode {Arrays} {Cultured} with in {Vitro} {Neural} {Networks} for {Motion} {Control} {Tasks}: {Encoding} and {Decoding} {Progress} and {Advances}},'' \emph{\BIBforeignlanguage{en}{Microsyst. Nanoeng}}, vol.~11, no.~1, Nov. 2025.

\bibitem{levyEnhancementNeuralRepresentation2012}
O.~Levy, N.~E. Ziv, and S.~Marom, ``\BIBforeignlanguage{en}{Enhancement of {Neural} {Representation} {Capacity} by {Modular} {Architecture} in {Networks} of {Cortical} {Neurons}},'' \emph{\BIBforeignlanguage{en}{Eur. J. Neurosci}}, vol.~35, no.~11, pp. 1753--1760, Apr. 2012.

\bibitem{forroModularMicrostructureDesign2018}
C.~Forró \emph{et~al.}, ``\BIBforeignlanguage{eng}{Modular {Microstructure} {Design} to {Build} {Neuronal} {Networks} of {Defined} {Functional} {Connectivity}},'' \emph{\BIBforeignlanguage{eng}{Biosens. Bioelectron}}, vol. 122, pp. 75--87, Dec. 2018.

\end{thebibliography}

\end{document}